\documentclass{aa}
\usepackage{graphicx}
\usepackage{rotating}

\begin{document}

   \title{Young Star Cluster Complexes
   in NGC 4038/39: Integral field spectroscopy using VIMOS-VLT 
   \thanks{Based on Observations at the Very Large Telescope of the
   European Southern Observatory, Paranal/Chile under Program
   71.B-3011}} 

   \titlerunning{Cluster Complexes in NGC~4038/39}

   \author{N. Bastian$^{1,2}$, E. Emsellem$^{3}$, M. Kissler-Patig$^{2}$,
         and C. Maraston$^{4}$
          }
   \authorrunning{Bastian et al.}
   \offprints{N. Bastian (bastian@star.ucl.ac.uk) \\ Current address:
         Department of Physics and Astronomy, University College
         London, Gower Street, London, WC1E 6BT}

   \institute{$^1$ Astronomical Institute, Utrecht University, 
              Princetonplein 5, NL-3584 CC Utrecht, The Netherlands \\
	      $^2$ European Southern Observatory, Karl-Schwarzchild-Strasse 2 
               D-85748 Garching b. M\"{u}nchen, Germany \\
	      $^3$ CRAL--Observatoire de Lyon, 9 avenue Charles
              Andr\'{e}, 69561 Saint-Genis-Laval Cedex, France\\
	      $^4$ University of Oxford, Denys Wilkinson Building,
              Keble Road, Oxford, OX13RH, United Kingdom
}

   \date{Received ???, 2005; accepted ??, 2005}


   \abstract{We present the first results of a survey to obtain {\it
   Integral Field Spectroscopy} of merging galaxies along the
   Toomre Sequence.  In the present work, we concentrate on the star
   cluster complexes in the Antennae galaxies (NGC~4038/39) in the
   overlap region as 
   well as the nuclear region of NGC 4038.  Using optical
   spectroscopy we derive the extinction,
   age, metallicity, velocity, velocity dispersion of the gas and star
   formation rate for each of the eight 
   complexes detected.  We 
   supplement this study with archival {\it HST-WFPC2 U, B, V,
   H$\alpha$, {\rm and} I} band imaging.  Correcting the observed
   colours of the star clusters within the complexes for
   extinction, measured through our optical spectra, we compare the
   clusters with simple stellar population models, with which we find an
   excellent agreement, and hence proceed to derive the ages and
   masses of the clusters from comparison with the
   models.  In five of the complexes we detect strong Wolf-Rayet
   emission features, indicating young ages (3--5~Myr). The ionized
   gas surrounding the complexes is expanding at speeds of
   $20-40$~km/s.  This slow expansion can be understood as a bubble,
   caused by the stellar winds and supernovae within  the complexes,
   expanding into the remnant of the progenitor giant molecular cloud.
   We also find that the complexes themselves are grouped, at about the
   largest scale of which young star clusters are correlated,
   representing the largest coherent star forming region.  We show
   that the area normalized star formation rates of
   these complexes clearly place them in the regime of star forming
   regions in starburst
   galaxies, thereby justifying the label of {\it localized
   starbursts}.  Finally, we estimate the stability
   of the complexes, and find that 
   they will probably loose a large fraction of their mass to the
   surrounding environment, although the central regions may merge
   into a single large star cluster.
    }
         
\maketitle

%

\section{Introduction}

Mergers of gas-rich spiral galaxies can produce vast amounts of young
massive star clusters.  Examples of these extraordinary systems are NGC
7252 (Miller et al.~1997, Schweizer \& Seitzer 1998), NGC 3921
(Schweizer et al.~1996), NGC 1316 (Goudfrooij et al.~2001) and NGC 3256
(Zepf et al.~1999), all of which harbour populations of  globular
cluster sized star clusters which formed concurrently with the merger.
The closest, and most spectacular ongoing spiral-spiral merger is NGC
4038/39 (also known as The Antennae) which 
contains thousands of massive star clusters in the process of forming
(Whitmore et al.~1999). Due to its proximity and high cluster
formation rate, the cluster population of NGC 4038/39 is an almost ideal
target to study cluster formation in its various stages as well as the
impact that the formation has on the surrounding ISM.  Additionally,
Whitmore et al.~(1999), among others, have noted several 'knots', or
cluster complexes, within the
disks of these galaxies, which are collections of many individual
young star clusters. 

Using a wealth of multi-wavelength data, Zhang et al.~(2001)
showed that the youngest star clusters in NGC 4038/39 are not
isolated, but instead tend to be distributed in a clustered fashion
themselves. This implies that clusters do not form in isolation, but
as part of a larger hierarchy of structure formation, extending from
stellar to kiloparsec scales (Elmegreen \& Efremov 1996).  These
complexes of star clusters are the subject of the present study.

The best studied cluster complex is a large complex in the spiral galaxy
NGC~6946.  This 
complex has a diameter of 600 pc and a total mass estimated to be $\sim
10^{7} M_{\odot}$.
The most massive star cluster within the complex has a mass of
$\sim10^{6} M_{\odot}$ and an age of 15 Myr (Larsen et al.~2002).
Additionally there are other smaller star clusters within the complex
which have similar ages (Elmegreen, Efremov \& Larsen 2000).  This
complex seems to have undergone three distinct bursts in its star
formation history, one between 20 and 30 Myr ago, the second, including
the most massive cluster, around 15 Myr ago, and a third which began
roughly 6 to 8 Myr ago and is continuing until the present (Larsen et
al.~2002).  Despite the multiple star formation epochs there is no
clear relation between the age of the star clusters and their
locations within the complex.

Other known large complexes (with and without star clusters) include the
Gould Belt in the Galaxy (e.g. Stothers \& Frogel 1974), a large
stellar complex in M83 
(Camer\'{o}n 2001), as well as a series of cluster complexes along
the western spiral arm of M~51.  These latter complexes seem to follow
the same mass vs. radius relation as giant molecular clouds, which appears
absent at the level of the individual star clusters (Bastian et
al. 2005b).  This suggests that these complexes are the direct results
of fragmentation of single giant molecular clouds. 

Kroupa (1998) and Fellhauer \& Kroupa (2002) have suggested that
cluster complexes (what they call superclusters) host a significant amount of
merging of clusters within them.  In their simulations this merging
results in a massive extended star cluster which shows a radial age and
metallicity gradient.  The authors suggest (Fellhauer \& Kroupa~2005)
that this is a mechanism to 
form extremely massive star clusters, like those observed in the
merger remnants, such as W3 in  NGC~7252 whose dynamical mass is
$\sim8~\times10^7M_{\odot}$ (Maraston et al.~2004).  Although their
model fails to reproduce the precise properties of W3 (they
under-predict the velocity dispersion) this mechanism may still be
valid for other massive clusters.

To better understand the nature of cluster and cluster complex
formation, we have obtained integral field spectroscopy of two
regions in the merging galaxies NGC 4038/39.  The intrinsic benefit of
integral field unit spectroscopy is that it removes the bias of a
pre-determined
axis of study, as is the case with classical long-slit spectroscopy.
This spectroscopic data is then combined 
with archival {\it HST-WFPC2} broad and narrow band imaging
which is able to spatially resolve each complex into its individual
cluster components.  The spectroscopic data presented here is one part of
a much larger survey of the relationship between young massive star
clusters, the interstellar medium, and the background stellar
populations in interacting galaxies.

In \S~\ref{obs} we present both the spectroscopic and photometric
observations.  The spectral features are introduced in \S~\ref{osp}
and we exploit them in \S~\ref{eandz} to derive their extinctions and
metallicities. \S~\ref{results} is dedicated to deriving the ages of
the complexes through the individual clusters within the complexes as
well as by Wolf-Rayet features when available.  The interstellar
matter within the complexes, the star formation rates, and expanding
shells are the topic of \S~\ref{expansion}.  In \S~\ref{formation} we
discuss the
formation of the complexes and their relation to giant molecular
clouds.  The stability of the complexes is the subject of \S~\ref{stability}
and in \S~\ref{conclusions} we summarize the main results.

\section{Observations}
\label{obs}

\subsection{VLT - VIMOS}
Two fields in the interacting galaxies NGC 4038/39, containing young
massive star clusters, were observed with 
the {\it VIMOS} (Visual Multi-Object Spectrograph) Integral Field Unit on
the {\it VLT} during the nights of April 3rd, 4th, and 5th, 2003.  The
first field 
(hereafter Field~1) is located in the interaction region between the
two galaxies (Region E + F, Fig.~6a in Whitmore et al.~1999), while
the second field (hereafter Field~2) is centered 
on the nucleus of NGC 4038.  The positions of the two fields are shown
in Fig.~\ref{ifu-positions}.  The observations were taken with the blue arm in 
high-resolution mode, covering the wavelength range
4150\AA~to~6100\AA~with a spectral resolution of $\sim1.8$\AA~at the
position of H$\beta$. 
The spatial sampling of the IFU was 0.66$''$ covering 27$''$ by 27$''$
on the sky.  Each individual exposure was 1200s long, and we obtained 8
useful exposures resulting in a total exposure time for
each field of 9600 seconds.  We also obtained off-target pointings
in order to subtract the sky background.  These are not used in the
present work as we used neighbouring regions of the complexes
to estimate the background, including the underlying contributions
from both the sky and galaxies NGC 4038/39. 

\begin{figure}[!htb]
      \caption{{\bf Left:} {\it HST-WFPC2} F555W mosaic image of the
      Antennae. {\bf Right:} The reconstructed {\it VLT-VIMOS/IFU}
      images shown in the same orientation, of the two pointings.
      Field~1 is shown on the top, while Field~2 is the bottom inset.} 
         \label{ifu-positions}
\end{figure}

The data were reduced using the {\it VIMOS/IFU}
pipeline\footnote{The pipeline is available at
http://www.eso.org/observing/gasgano/vimos-pipe-recipes.html}.  The
pipeline includes bias subtraction, flat-fielding, extraction of
individual spectra, and wavelength calibration.  The pipeline also
produces reconstructed images of the field of view, which are shown
in Fig.~\ref{ifu-positions} (right).  To obtain the spectra of each cluster
complex, we summed the spectra from all the 'spaxels' (spatial
elements of the reconstructed image) containing flux from the cluster complex.
The apertures were selected from the reconstructed images, and
typically contained 12 to 18 spaxels.  Due to the irregular apertures
and the sizes of the complexes being comparable to the spatial
resolution, no attempt was made to resolve the individual complexes.
We then selected a region of background near each complex in
order to extract a background
spectrum, which was then subtracted from each cluster complex
spectrum. These spectra were then flux-calibrated using photometric
standard stars observed in each of the four {\it VIMOS/IFU} quadrants.

\subsection{HST - WFPC2}

We retrieved {\it HST WFPC2} observations in the F336W (U), F439W (B),
F555W (V), F658N (H$\alpha$), and F814W (I) filters from the {\it HST}
archive.  These data are presented in depth in Whitmore et
al.~(1999). Point-like sources were found using the {\it DAOFIND} task
in {\it IRAF} and aperture photometry was carried out with the {\it
PHOT} routine using an aperture, inner and outer background radii of
1.5, 3.5 and 5.5 respectively.  CTE corrections were carried out using
the formulae of  Whitmore, Heyer \& Casertano (1999).  Aperture
corrections were determined
from bright isolated sources.  Comparison of our photometry to that of
Whitmore et al.~(1999) reveals that they are consistent within 0.03 in
magnitude.
Figures~\ref{image-f1}~\&~\ref{image-f2} show the F439W {\it
HST-WFPC2} images of Field~1 and
Field~2 respectively.  We adopt the same distance to the Antennae as
Whitmore et al.~(1999), namely 19.2 Mpc.  At this distance, 1
arcsecond corresponds to 93~pc so one wide field {\it WFPC2} pixel and
one {\it VIMOS-IFU} spaxel correspond to 9.3~pc and 61~pc repsectively.

The complexes in Field~1 are named Complex 1 through 6, while the
complexes in Field~2 are named Nuc~1 and Nuc~2 to designate them as
belonging to the nuclear region of NGC~4038.

\begin{figure}[!h]
\begin{center}
      \caption{{\it HST-WFPC2} F439W image of the overlap region in
      NGC~4038.  The image is $\sim$2.5 kpc (27.5'') on a side.  The complexes are
      identified and labelled Complex~1 through 6.} 
         \label{image-f1}
	 \end{center}
\end{figure}


\begin{figure}[!h]
      \caption{{\bf Field~2.}  F439W (B) band image of the nuclear
      region of NGC 4038.  The two complexes in this region are
      labelled Nuc~1 and Nuc~2.  Each side of the image is
      $\sim$2.5~kpc (27.5'').}
         \label{image-f2}
\end{figure}

\section{Observed spectral properties}
\label{osp}

\subsection{Introduction to the spectra}

\begin{figure}[!h]
    \includegraphics[angle=0,width=9cm]{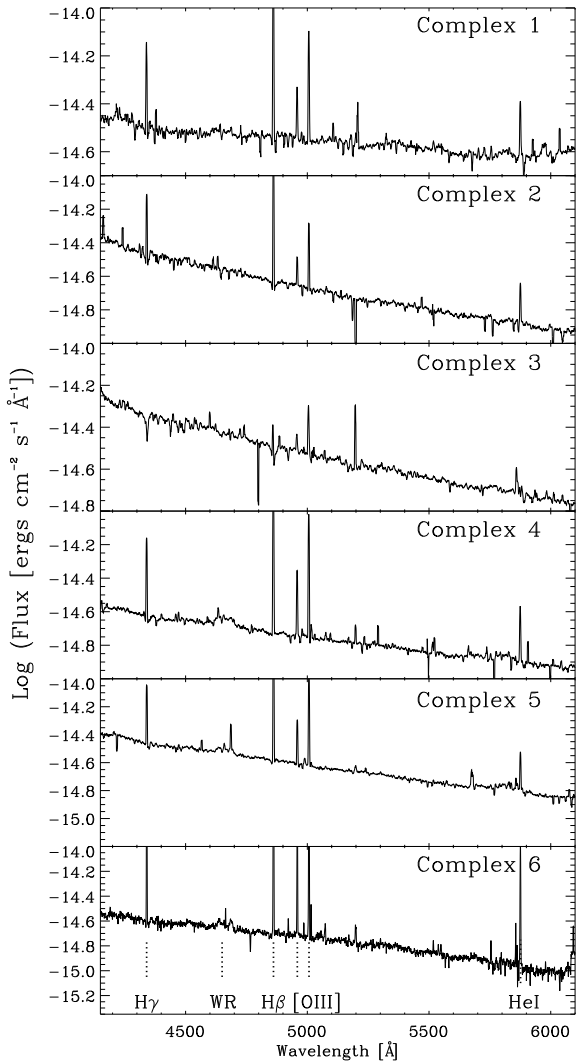}
    \caption{Observed spectra of the cluster complexes in Field~1.
    All spectra are in the rest wavelength of NGC 4038, and have not
    been corrected for extinction.  Prominent emission lines are
    indicated, including the Wolf-Rayet feature at $\sim 4650$ \AA.
    The spectra have been smoothed using a boxcar function with width
    4.5 \AA.}
         \label{spec-tot-f1}
\end{figure}

\begin{figure}[!h]
     \includegraphics[angle=0,width=9cm]{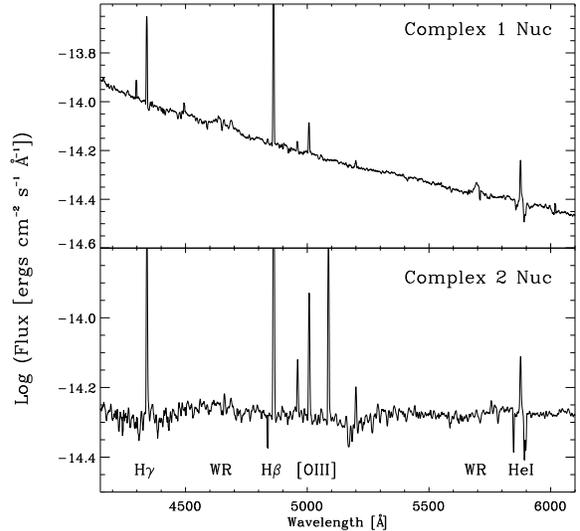}
    \caption{Observed spectra of the cluster complexes in Field~2.
    All spectra are in the rest wavelength of NGC 4038, and have not
    been corrected for extinction.  Prominent emission lines are
    indicated, including the Wolf-Rayet feature at $\sim 4650$ \AA~
    (the blue bump) as well as the Wolf-Rayet feature at 5696 \AA~
    which is the broad emission from C III.
    The spectra have been binned using a boxcar function with width
    4.5 \AA.}
         \label{spec-tot-f2}
\end{figure}

The full spectra of each of the complexes are shown in
Figs.~\ref{spec-tot-f1}~\&~\ref{spec-tot-f2}.  The dominating features
of these spectra are the nebular emission lines H$\gamma$, H$\beta$,
[O~III] ($\lambda\lambda4959,5007$\AA) and He~I ($\lambda5876$),
indicating the presence of a large amount of ionized gas.
Additionally, Complexes~4, 5, Nuc~1, and Nuc~2 show strong Wolf-Rayet
emission features (e.g. He II~$\lambda$4686), characteristic of
extremely young stellar populations.  Many of the complexes
also show underlying stellar absorption features, the most notable
example of this is Complex~3.  All of these features will be discussed
in detail in the following sections. 

\subsection{Nebular emission line features}

\begin{figure}[!h]
     \includegraphics[angle=0,width=9cm]{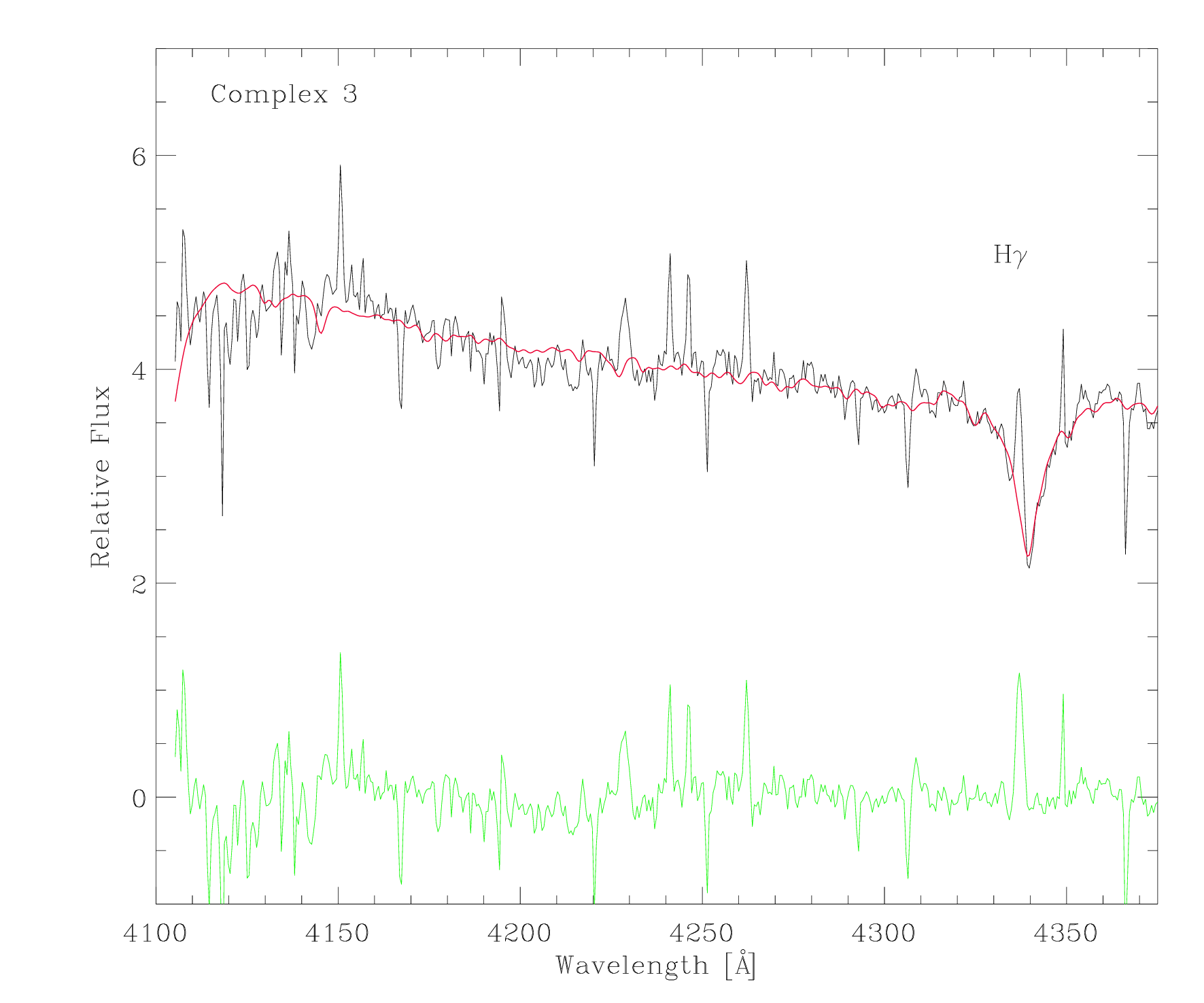}
     \includegraphics[angle=0,width=9cm]{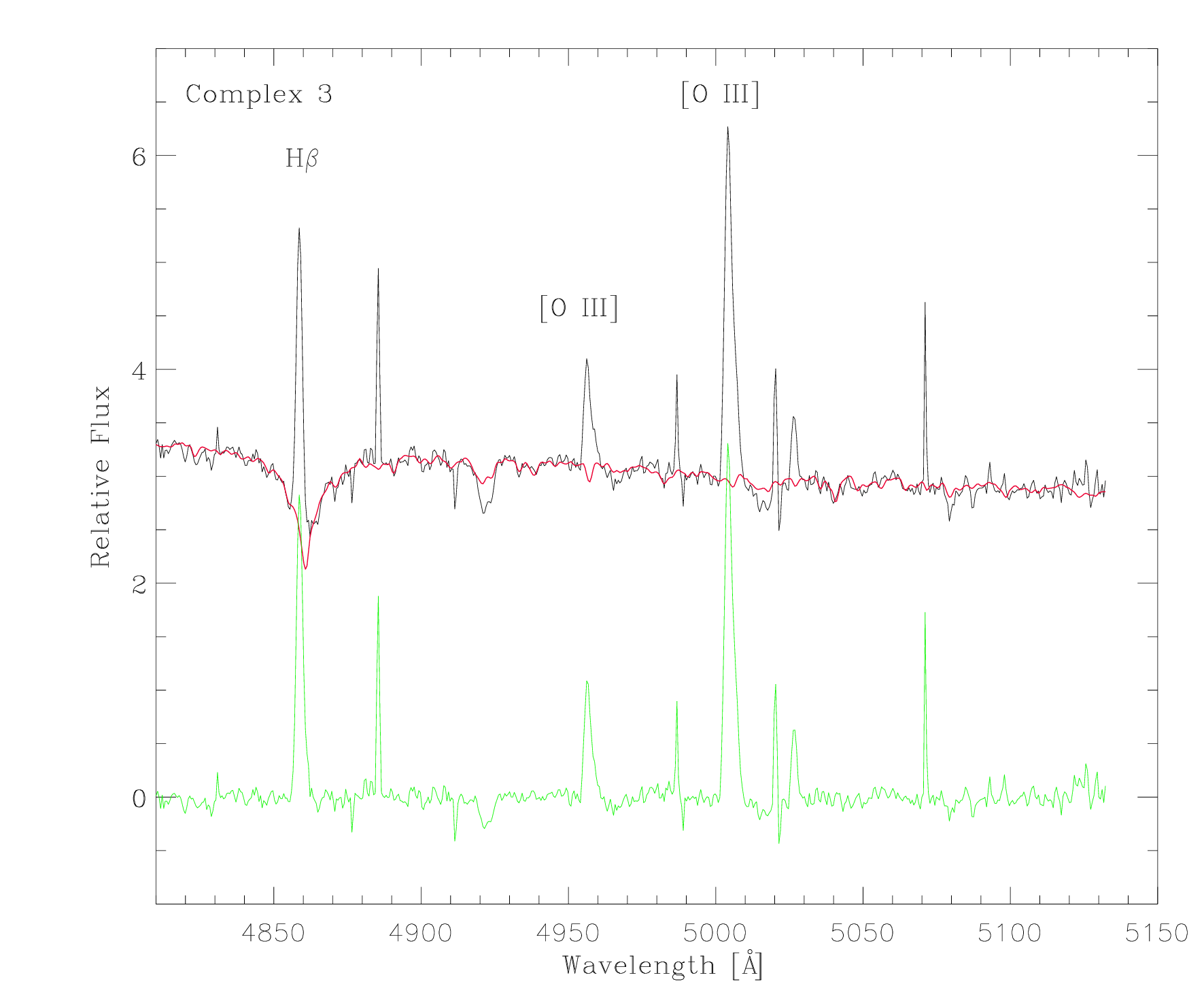}
 
      \caption{{\bf Top:} The blue region of the spectrum of
         Complex~3 in Field~1.  The black line is the observed
         spectrum, the red 
         line is the best fitting stellar template, and the green line
         is the pure emission line spectrum (the observed spectrum
         minus the stellar template).  {\bf Bottom:} The same as the
         top panel, but now for the red section of the spectrum of
         Complex~3.  } 
         \label{ppxfmethod}
\end{figure}

Many of the spectra of the cluster complexes show a combination of (nebular)
emission and underlying stellar absorption features.  In the present
study we are primarily interested in the emission features, namely
H$\gamma$, H$\beta$, and [O~III]$\lambda\lambda4959,5007$ which can be
used to estimate the extinction, metallicity, and total ionizing flux
in the region.  In order to remove the underlying absorption features
from our data, we have employed the penalized pixel fitting method
developed by Cappellari \& Emsellem (2004).  This procedure fits a
linear combination of template models (in this case the simple stellar
population models) with the addition of a polynomial, to
the observed spectra.  We use the Vazdekis (1999) simple stellar
population models as templates because these models have a spectral
resolution ($\sim1.8$\AA) that is comparable to that of the observed
spectra.   The best fitting combination
template was then subtracted from the observed spectra, leaving a pure
emission line spectrum.  An example of this technique for two sections
of the spectra of Complex~3 is shown in Fig.~\ref{ppxfmethod}.

We then measured the wavelength, flux, and FWHM of each of the emission
lines with the {\it SPLOT} routine in {\it IRAF}, approximating each
line as a Gaussian.  For the kinematic determination we measured the H$\gamma$,
H$\beta$, [O~III]$\lambda\lambda4959,5007$, and HeII($\lambda5876$)
emission lines
when available.  Typical velocity errors are $\sim5-10$~km/s,
calculated as the standard deviation of the velocities measured from
each individual line.
The derived extinctions and velocities are shown in
Table~\ref{table:2info}.  We note that our Complexes~1 in Field~1
corresponds to Region~E as well as Complexes~4 and 5 corresponding to
Region~F of Whitmore et al.~(2005).  Comparing the derived velocities
between that work and ours shows that the measured velocities differ
by less than 5~km/s.

We do not detect [O~III]$\lambda4363$ in any of our spectra, thus we
were not able to estimate the temperature of the ionized regions.
Non-detection of this line suggests metallicities $> 0.5$
$(O/H)_{\odot}$ (Kennicutt et al.~2003).  As will be seen in the next
section, this is consistent with what we find through the comparison
with stellar population models as well as from empirically calibrated
emission line ratios, with all complexes studied here being nearly or
above 0.5~Z$_{\odot}$.

\section{Extinction and metallicity of the complexes}
\label{eandz}

By measuring the relative strengths of H$\gamma$ to H$\beta$, we can
estimate the amount of extinction within the complex.  For Case~B
recombination at 10,000K the ratio of H$\gamma$ to H$\beta$ is
0.466 (Osterbrock 1989).  By comparing the observed value of H$\gamma$/H$\beta$ and
assuming the extinction curve of Savage \& Mathis (1979) we find
extinctions, A$_{\rm F555W}$, between 0 and $\sim1.5$.  Care must be taken
when interpreting these numbers, however, as extinction most likely
varies across each complex.  Unfortunately our observations do not
have the spatial resolution to correct for this effect.  We therefore
apply the average extinction estimate of 
each complex to each source within that complex.  The one exception to
this is Complex~1.  The lack of $H\alpha$ emission at the position of Complex~1
and the strong HII 
region nearby lead us to conclude that the emission lines and
Wolf-Rayet features come from the neighboring HII region, and not
from Complex~1 itself.  Thus, we do not know the extinction for
Complex~1 and we therefore assign the clusters within Complex~1 zero
extinction. 

We can also use the spectra to estimate the metallicity of each
 complex.  As is the case with extinction, we do not have high enough
 spatial resolution to detect variations in metallicity across the
 complex.  However, we do not expect large metallicity gradients
 within each complex as all sources within a complex presumably formed
 from the same giant molecular cloud or cloud complex.  From Vacca \&
 Conti (1992), we derive the O/H ratio in the emission line regions using: 

$$ {\rm log} ({\rm O/H}) = -0.69 {\rm log} R_{3} - 3.24 (-0.6 \leq {\rm
log} R_{3} \leq 1.0) $$

where 

$$ R_{3}=\frac{I([{\rm O~III}]\lambda4959) + I([{\rm
O~III}]\lambda5007)}{I(\rm{H}\beta)} $$

The estimated intrinsic uncertainty in the calibration of this method
is $\pm$ 0.2 in log (O/H) (Edmunds \& Pagel 1984).  We have adopted
the ``Local Galactic'' oxygen abundance of (O/H)$_{\odot} = 8.30
\times 10^{-4}$(Meyer 1985) for the ``solar'' abundance. 

This empirical relation, however, has been shown to be affected
by the hardness of the ionizing photons.  Another empirical relation
has been introduced which attempts to compensate for this effect, the
R$_{23}$ relation. The relation includes the contribution from 
another ionized species of oxygen, namely [O~II], which takes into
account the ionization structure of oxygen.  Unfortunately our wavelength
range does not extend to the [O~II]$\lambda3727$\AA~line, which is required for
the R$_{23}$ calibration.  Also see 
Kennicutt et al.~(2003) for a discussion on the accuracy of using
collisionally excited lines for metallicity determination.

Despite the above caveats, we list the measured metallicities of the
complexes in Table~\ref{table:2info}.

\begin{figure}[!h]

\hspace{-0.5cm}    
      \caption{Reconstructed image of the total intensity of the {\it
         VIMOS-IFU} observations
         of Field~1 with the five complexes labeled by number (the
         sixth is out of the field of view on many of the expssures so
         we will not study the spectrum in detail in the present work).  The
         orientation is the same as in Fig.~\ref{image-f1}, and the
         image is 27'' on a side ($\sim2.5$~kpc).}  
         \label{image-vlt-f1}

\end{figure}

\section{Ages of the cluster complexes}
\label{results}

\subsection{Ages of the individual clusters based on photometry}
\label{agesphot}

A common technique employed to derive the ages of individual clusters
is to compare the spectral energy distribution of each cluster to that of
simple stellar population models (e.g. Bastian et
al.~2005a).  We use simple stellar population models (SSP) from Maraston
(2005).  At ages $<$ 30 Myr, these models are based on the
tracks/isochrones from the Geneva group (Schaller et al.~1992,
Schaller et al.~1993) and span the metallicity Z/H range from 1/50 to
2~$Z_{\odot}$.  The IMF can be Salpeter (1955) or Kroupa (2003), from
0.8 to 120~$M_{\odot}$ (where we have chosen the Kroupa type IMF).

\begin{figure*}[!htb]
\begin{centering}
      \includegraphics[angle=0,width=12cm]{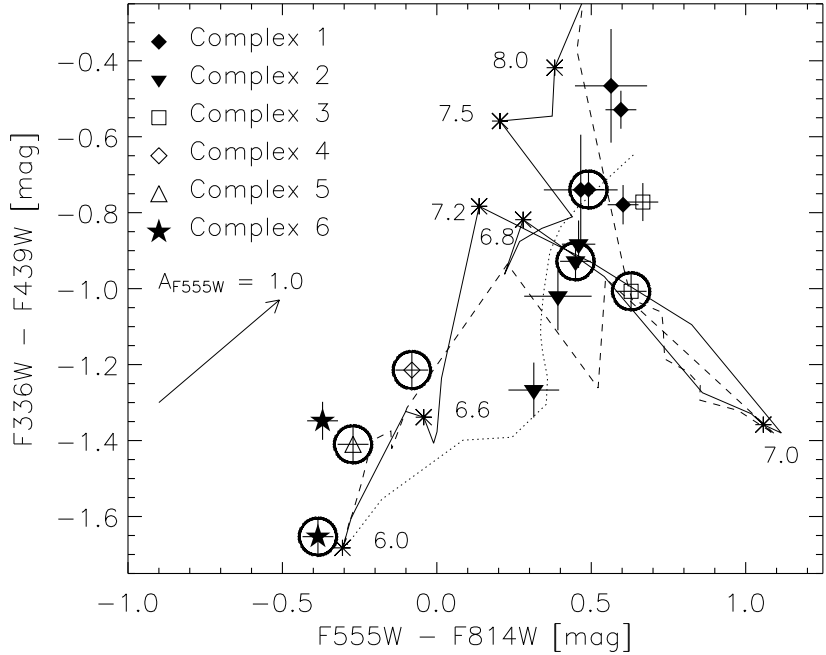}
    \caption{Colour - colour diagram of the clusters within
         the five complexes in Field~1. The dashed and solid
         lines are the Maraston model (2005) tracks for
         solar and twice solar metallicity,
         respectively (the numbers correspond to the logarithm of the
         age in years at that point).  The dotted line represents
    the changing colours of a model of continuous star formation of
    solar metallicity (Maraston 2005), with the final point
    representing an age of 1~Gyr. All points have been corrected for
    the average
         extinction of the complex (see text).  
         The brightest cluster in each complex is circled.  Sources
         within each complex have been labeled as follows; solid
         diamonds for Complex~1, filled  triangles for
         Complex~2, open squares for Complex~3, open diamonds for
         Complex~4, and open triangles for Complex~5 } 
         \label{colour-f1}
	 \end{centering}
\end{figure*}

One of the main degeneracies in this kind of study, is between age
and extinction for young clusters.  This can be seen in
Fig.~\ref{colour-f1} where the extinction vector is almost parallel to
the SSP models for much of the first $\sim10^{7.5}$ years of
a cluster's existence .  In
the present study, however, we estimated the extinction
independently, using the H$\gamma$/H$\beta$ ratio from our
spectroscopy. Therefore, we have corrected the photometry of each
cluster within the complexes for the extinction value derived in
\S~\ref{eandz} (except Complex~1 as discussed above).   It should
be noted that the extinction for the individual clusters is
not necessarily the same as the nebular extinction (as derived from
the H$\gamma$/H$\beta$ ratio).  However, if the emission lines are
generated near the clusters and the majority of the extinction occurs
outside this region, then the extinction of the clusters and the
nebular extinction should be approximately equal.  Throughout this
work we will assume that this assumption holds.

\begin{figure}
\begin{centering} 
    \includegraphics[angle=0,width=9cm]{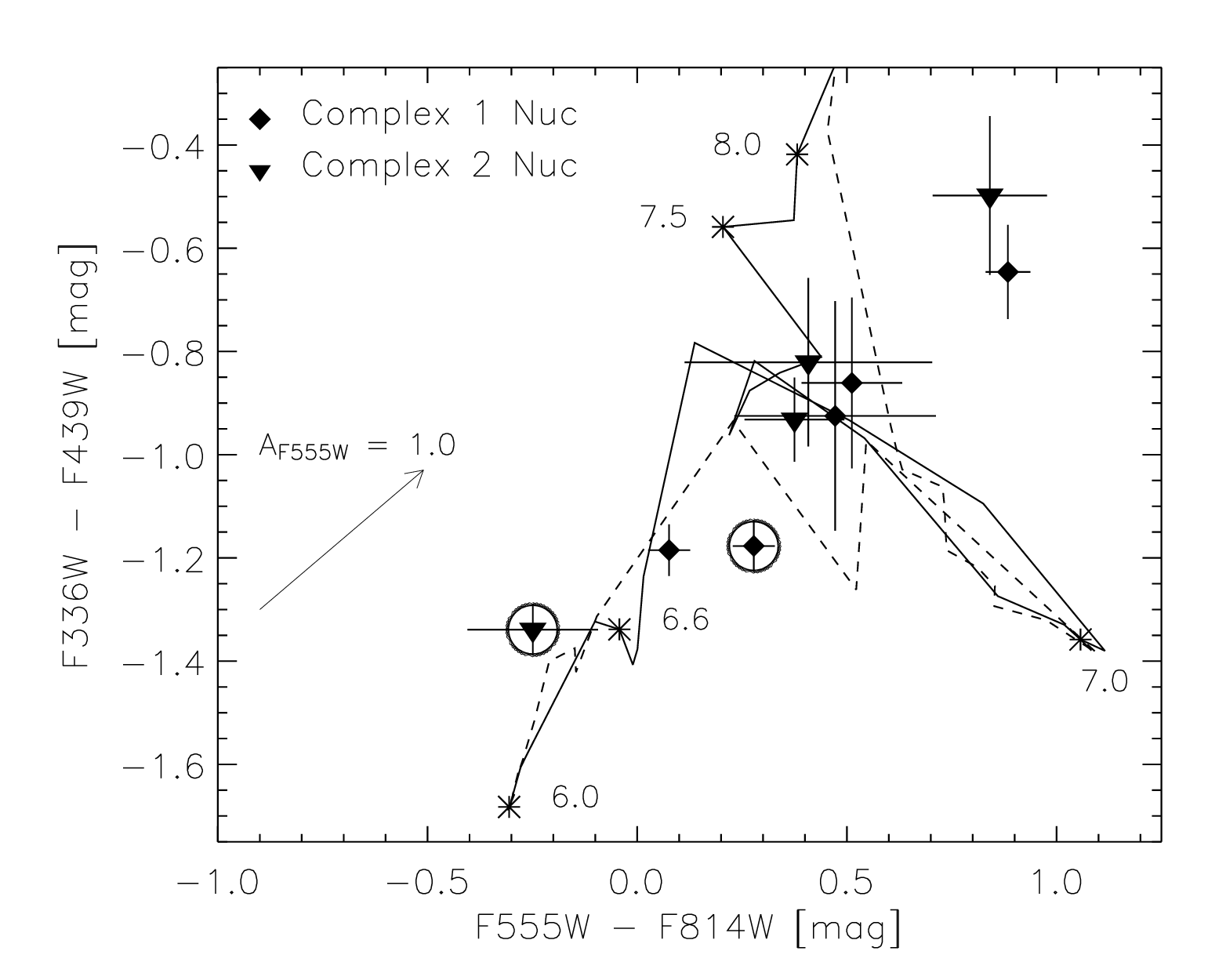}
      \caption{Same as Fig.~\ref{colour-f1} but now for clusters
      inside the complexes in the nuclear region of NGC 4038
      (Field~2).  Filled diamonds are clusters within Complex~Nuc~1
      while 
      filled triangles represent clusters in Complex~Nuc~2.  The two
      clusters from complex Nuc~1 near the lower left model tracks
      dominate the optical light from the complex.  Complex~Nuc~2
      seems to have clusters within its projected radius with ages
      between a few Myr and a few tens of Myr, although we note that
      the clusters lie parallel to the extinction vector
      suggesting a similar age but variable extinction.  The brightest
      cluster in 
      each complex is circled.}
         \label{colour-f2}
\end{centering}
\end{figure}

Figures~\ref{colour-f1}~\&~\ref{colour-f2} show the colours of all
clusters that have errors of less than 0.15 in their colours. The brightest
cluster in each complex is circled.  The Maraston (2005)
SSP models of different metallicities,  $Z_{\odot}$, $2
Z_{\odot}$ are overplotted as 
solid and dashed lines respectively.   The dotted line in
Fig.~\ref{colour-f1} represents a solar metallicity continous star
formation model.  The numbers
correspond to the log of the age of the
models (in years), which are also marked with an asterisk.

The models do an excellent
job in reproducing the observed colours, once extinction is
removed. The masses of the complexes are derived by measuring the
light of the complex (corrected for extinction) and comparing it with
the mass-to-light ratios (including remnants) of the SSP models. 
Table~\ref{age:table} shows the derived ages and masses from the
photometric method.  The age (and hence the mass) of Complex~3 is
poorly constrained as it is located near the 'red loop' in colour space
for solar metallicity SSP models. 

The photometric ages of the brightest cluster in each complex derived
in this section are shown in Table~\ref{age:table}, and 
range from a few Myr to $\sim25$ Myr.  We note that the relative ages
of the brightest clusters in Complexes~4, 5, and 6, are consistent
with the size of the HII region surrounding each of them, in the sense
that the youngest complex (Complex~6) has a smaller HII region
surrounding it.  Presumably, the bubble caused by the winds and
supernovae inside this complex has not had enough time to expand to a
larger extent.  We will return to this point in Sec.~\ref{expansionvelocities}.

From the current data (see
Figs.~\ref{colour-f1}~\&~\ref{colour-f2}) 
we cannot rule out the existence of clusters with different ages
($\Delta({\rm age)}~\geq~$5~Myr)
within the same complex.  This is due to the colours of different
clusters lying parellel to the extinction vector.  Spectroscopy of
individual clusters is required to derive the extinction of each
individual cluster in order to break the age/extinction degeneracy.
We may expect an age difference within a single complex based on 
the works of Elmegreen et al.~(2000) and Larsen et
al.~(2002) who found clusters with different ages in the large cluster
complex in NGC 6946 ($\Delta$(age)$\approx 30$ Myr).  Despite the possible
presence of these multi-age 
populations, we assign the age of the brightest cluster to that of the
complex for further comparison.

We can most clearly see the caveat of our extinction determination in Complex~Nuc~2
in Fig.~\ref{colour-f2}.  There, we see that all the clusters
within the complex can be connected by the extinction arrow in colour
space.  Thus, these clusters could be older, low extinction objects,
or younger, high extinction objects.  So we can only conclude that
there is either a large difference in reddening across the complex or
that there exists a multiple age population.  Spectroscopy of
individual complex members is required to break this degeneracy.

\subsection{Age dating of the complexes based on their Wolf-Rayet features}
\label{ageswr}

\begin{figure}[!h]
     \includegraphics[angle=0,width=8cm]{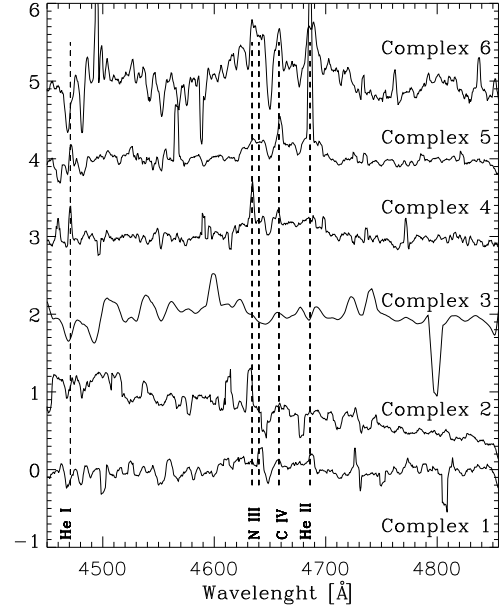}
      \caption{The region of the spectrum around the 'blue bump'
      Wolf-Rayet ($\sim\lambda4650$\AA, made up of the N~III, C~IV and
      He~II broad emmision lines) feature for the five complexes in Field~1.
      The continuum has been subtracted from each spectrum.
      The WR feature is clearly observed in complexes 4 and 5, and
      weakly, though definitely present, in complex 1.}
         \label{wrspectra-f1}
\end{figure}

\begin{figure}[!h]
     \includegraphics[angle=0,width=8cm]{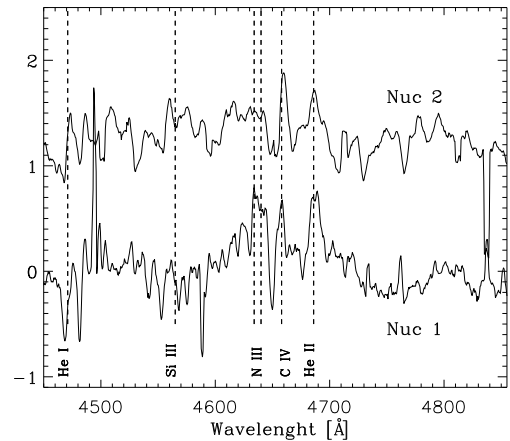}
      \caption{Same as Fig.~\ref{wrspectra-f1} but now for the two
      complexes present in the nuclear region of NGC 4038.  Complex~Nuc~1 has a strong WR feature, while in Nuc~2 the feature is
      either weak or absent.}
         \label{wrspectra-f2}
\end{figure}

We can take advantage of the presence of young massive stars within each
complex by using their Wolf-Rayet features to independently derive their
ages. Wolf-Rayet stars produce strong broad emission lines which can
be identified in the composite spectrum of a stellar
population.  This technique has the advantage that it is independent of
extinction and can be used to verify the ages determined
through their colours.  The spectral region around the
``blue bump'' Wolf-Rayet feature ($\lambda4650$\AA) for each of the
complexes is shown in Figs.~\ref{wrspectra-f1}~\&~\ref{wrspectra-f2}.

The spatial resolution of the observations allow us 
to find where strong Wolf-Rayet features are present within the field
of view.  To do this, we fit a polynomial to the spectra region
around the 'blue bump' (see Fig.~\ref{wrspectra-f1}),
and then subtract the continuum.  We then summed the remaining flux in
that region, and reconstruct the spatial image.  The results of this
are shown in Fig.~\ref{image-vlt-wr} where bright colours represent
strong Wolf-Rayet features and dark colours represent regions where
little or no features are detected.  Here we see that
Complexes~4, 5, and 6 (Complex 6 being mostly off the image however)
are the only regions of strong WR features in this field.  From the fact
that these three complexes show WR features, we can conclude that they
(or at least the massive stars within the complex) have the same age
within $\sim4$ Myr (Norris 2003), and all have ages less than 10 Myr.
This strongly suggests that the formation of these complexes, or at
least the most recent round of star formation within them, was
triggered at about the same time.  We shall return to this point in
Sec.~\ref{formation}.  We also point out that Complexes~4, 5,
\&~6 do not appear to have any old clusters within them (see
Fig~\ref{colour-f1}) suggesting that the most recently round of star
formation is the only significant period of star formation to
take place within these complexes.  While we cannot rule out the
presence of an older population of stars within the complexes (age
$>$~100~Myr), we do note that such a population would not affect
colours, and hence the derived ages, of the detected clusters in these
complexes due to the fact that young clusters significantly outshine
their older counterparts in the optical wavelengths.

Finally, we note that the clusters in
Complexes~4, 5, \&~6 lie far from the continuous star formation models
(in colour space) and much closer to the SSP model colours, whereas
the clusters in Complex~2 are also compatible with the continous
cluster formation models (the dotted line in Fig.~\ref{colour-f1}).

\begin{figure}[!tb]

\hspace{-0.5cm} 
      \caption{The spatial distribution of Wolf-Rayet features in
         Field~1.  Brighter colours indicates strong Wolf-Rayet
         features.  Note that Complexes~4, 5, 6 all show strong WR
         features.  The
         orientation is the same as in Fig.~\ref{image-f1}.   One
         spaxel (spatial pixel) in this image is 63~pc.}  
         \label{image-vlt-wr}

\end{figure}

\section{Interstellar matter in the cluster complexes}
\label{expansion}

\subsection{General velocity field}

 Figure~\ref{velocity-field1} shows the velocity field of the gas (fit
on both the H$\beta$ and [OIII]$\lambda\lambda4959,5007$\AA~absorption
lines) in
Field~1.  The complexes and their positions are indicated.  The scale
runs from 1400~km/s (blue) to 1650~km/s (white).  In general we see
that the velocity changes from higher to lower from the bottom of the
image to the top.  This is in good agreement with the general velocity
pattern of the Antennae in this region, i.e. as measured in HI
(Hibbard et al.~2001).

\begin{figure}[!h]
      \caption{The velocity field of Field 1.  The scale runs from
      1400~km/s (blue) to 1650~km/s (white).   One
         spaxel (spatial pixel) in this image is 63~pc.}  
         \label{velocity-field1}
\end{figure}

\subsection{Expansion velocities}
\label{expansionvelocities}

 In order to estimate the velocity dispersion of the gas (a
 measure of the random motions of the gas), we have
fit the width of the H$\beta$ and
 [O~III]$\lambda\lambda4959,5007$  emission lines after
subtraction of the best-fitting stellar background template. 
To correct for
the instrumental resolution, we also measured the sky line at
$\lambda5577$ and subtracted the width of this line in quadrature for
the H$\beta$/[O~III] combination.  This routine was carried out for
every spectrum (i.e. for every spaxel) in our data cube.  The
resulting velocity dispersion 
map for Field~1 is shown in Fig.~\ref{image-vlt-sigma}, where dark
blue represents the regions with the lowest velocity dispersion and
white represents regions with the highest dispersion (the scale runs
from 0 to 100 km/s).  The average value for each complexe is shown in Table~\ref{table:2info}.

The complexes (marked with circles and labeled in the figure) and the
regions around them, show small velocity dispersions, of the order of
20-40~km/s. These values can be understood as bubbles, originating
from a cluster wind, which is made up from the collection of winds
from hot stars and supernovae.  If one assumes that the winds/SNe act
collectively to form a cluster wind, then one can estimate the speed
of expansion of the resulting bubble using equations 12.12 and 12.19
in Lamers \& Cassinelli (1999).  For this we assume that the ambient
medium that the bubble is expanding into has a typical density of a
GMC in the Antennae (0.3 $M_{\odot}$/pc$^3$ - Wilson et al.~(2003))
and a wind kinetic luminosity of between $10^{39}$ and
$10^{42}$~ergs/s (estimated from Tenorio-Tagle et al.~2005).  We
further assume that the wind has been blowing 
constantly for the past 3-7~Myr.  With these assumptions (and assuming
that the driven bubble is in the energy or momentum 'snowplow' phase)
the bubble outflow velocity is expected to be between 10 and 35 km/s.
These values are in excellent agreement with the observed velocity
dispersions.

The exception to the above argument is Complex~3, which has a much
higher velocity dispersion than the other complexes.  However, we do
not detect strong H$\alpha$ or H$\beta$ emission associated with
this complex (the sources within Complex~3 also appear older than 10
Myr based on their colours), hence the velocity dispersion that we
measure is likely to be associated with the background.

\subsection{Star formation rates}

We can also estimate the star formation rates of the complexes.  Our
optical spectra do not extend to the H$\alpha$ emission line,
therefore we will use the dereddened H$\beta$ line.   To do this, we
measure the flux of the H$\beta$ line, correct for
extinction (as measured in \S~\ref{eandz}) and finally convert this flux to
the intrinsic H$\alpha$ flux using the intrinsic line ratios.  We assume
Case B recombination (10,000 K) giving a H$\alpha$ to H$\beta$ ratio
of 2.85.  We then assume a star formation rate proportional to the
H$\alpha$ flux as $\Sigma_{\rm SFR}(M_{\odot}~{\rm yr}^{-1}) = 7.9
\cdot 10^{-42} L(H\alpha)({\rm ergs~s}^{-1})$ (Kennicutt 1998).  
The measured star formation rates are given in Table~\ref{age:table}.
Again, we note that the emission features seen
in Complex~1 are likely due to the overlapping HII region.  

The star formation rates of the complexes are between 0.2 and $\sim
1.4 M_{\odot}$ yr$^{-1}$.  These rates are extremely high, considering
their relatively small sizes.  Using the radius determined in the
F555W band (and assuming that the complexes are circular) the star
formation rate per kpc$^2$ ($M_{\odot}$~${\rm yr}^{-1}$~${\rm kpc}^{-2}$)
is between 1 and 10.  This is comparable to circumnuclear starbursts
(Kennicutt 1998) and more than 10 times higher than the definition for
a starburst put forward by Kennicutt et al.~(2004).

These high star formation rates show
that the complexes studied here can be thought of as {\it localized
starbursts} (terminology from Efremov 2004).  This coherent and intense
form of star formation will significantly affect the surrounding ISM,
both in distributing metals and in adding turbulence.  These localized
starbursts are expected
to create a patchy metal distribution in the ISM, as observed in the 
Antennae (Fabbiano et al.~2004).

\begin{figure}[!h]

\hspace{-0.5cm}
      \caption{The velocity dispersion of the gas as measured from the
         combination of H$\beta$ and the
         [O~III]$\lambda\lambda4959,5007$ lines.  The colour coding
         extends from 0 to 100 km/s, with blue representing the lowest
         velocity dispersions, and white representing the highest.
         The measurements have been corrected by subtracting in
         quadrature the measured width of the skyline at
         $\lambda5577$.  Note that the gas at the position of the
         complexes has the lowest velocity dispersion in the region
         (20-40~km/s).   One
         spaxel (spatial pixel) in this image is 63~pc.}  
         \label{image-vlt-sigma}

\end{figure}

\section{Formation of the complexes}
\label{formation}

\subsection{Insights from the HII}

Figure~\ref{source-ha} shows the {\it HST-WFPC2} continuum subtracted
H$\alpha$ images for 5 of the complexes in our sample which have
extended H$\alpha$ emission.  The images are centered on the brightest
cluster within each complex, which are marked with a red cross.
Additionally, in two of the complexes (Complex~2 and Complex~Nuc~1) we
indicate another cluster within the complex which appears to be the
source of the ionization.  The emission in Complex~6 (not shown) is
concentrated on the brightest cluster, presumably due to the clusters
very young age, which has not had time to create an extended ionization bubble.

In Complex~2, the brightest source appears to be older than other
parts of the complex, and not responsible for the HII region. There is
another bright source, which appears to be very young (6--9 Myr from
Fig.~\ref{colour-f1}). This bright young source is located near the
center of the ionized region, suggesting that it is responsible for
it.  The photometry of the brightest source in Complex~2 gives an age
of $\sim25$ Myr.  At this age, no significant H$\alpha$ emission is
expected.    If,
on the other hand, the brightest cluster in this complex has a larger
extinction than estimated in \S~\ref{eandz} then we would expect to
see H$\alpha$ emission, which is not present. The interpretation of
this is that this complex contains clusters with different ages.
This is highly reminiscent of the star cluster Hodge~301 in 30
Doradus.  This cluster has an age of $\sim25$ Myr, much older than
R136 (Grebel \& Chu 2000).
Therefore multiple episodes of cluster formation within complexes may
be a common feature.

As may have been expected, the complexes near the nucleus of NGC 4038
appear as the most complicated.  Complex~Nuc~2 is made up of multiple ionized
regions, whereas Complex~Nuc~1 also has a clear bubble like
structure.  The center of this bubble seems completely devoid of
$H\alpha$ and any optical source.  The eastern region of the bubble
contains two bright young clusters which may be the ionizing sources.

\begin{figure}[!hbt]
      \caption{The positions of the brightest cluster within each
      complex, superimposed on a continuum subtracted H$\alpha$
      image. The complex IDs are given in each panel.  Each image is
      centered on the brightest source, other 
      sources shown are other young clusters in the complex as seen
      from Figures~\ref{colour-f1}~\&~\ref{colour-f2}, presumably
      responsible for creating the surrounding HII regions. All panels have the
      same scale, Complexes~2, 4 and 5 share the same orientation,
      while Nuc~1 and Nuc~2 also have the same orientation.}  
         \label{source-ha}
\end{figure}

\subsection{Groupings of Complexes}
In Fig.~\ref{halpha-contours-f1} we show the continuum subtracted {\it
HST-WFPC2} H$\alpha$ image, along with the intensity contours.  The
sizes of the complexes in H$\alpha$ and 
the F555W bands is given in Table~\ref{age:table}.  The complexes are
roughly the typical size of GMCs, and the masses derived are similar
to that expected for GMCs of similar sizes\footnote{Using the data of
spatially resolved clouds in The Antennae from Wilson et al.~(2003),
we find that $M_{\rm vir} = 0.024 \cdot R^{3.47}$, where M and R are
in units of M$_{\odot}$ and pc respectively.}.  The large ionized
region surrounding Complexes~4, 5,and 6
has approximately the spatial size ($\sim735$~pc) where the two-point
correlation
function of young clusters drops suddenly (Zhang et al.~2001).  This
is also approximately the size of large GMCs or GMC complexes within
the Antennae (Wilson et al.~2003).  The common H$\alpha$ envelope
around this grouping of complexes and the presence of Wolf-Rayet
features in each of the complexes, suggests that all of these complexes
began to form at roughly the same time ($\Delta$(age) $<$ 5-6 Myr).
Complexes~4 and 5 also have extremely similar metallicities.  Thus
this grouping of complexes may have formed from a single GMC (or GMC
complex) with multiple massive cores.   Again we note that the
presence of WR features and common H~II envelope in Complexes~4, 5,
\&~6 only suggests that the most recent episode of star formation was
coeval.  But the lack of older clusters in these complexes suggests
that the most recent episode of star formation was also the only or
most significant bout of star formation. 

If we assume that this grouping of complexes formed out of the same
GMC or GMC association, then it is interesting to compare the
formation timescale and the sound-travel time.  Complexes~4 and 5 are
separated by $\sim300$~pc, while Complexes~5 and 6 are $\sim350$~pc
(note however that these are lower limits as these are projected
distances).  Assuming that the complexes have
ages which are the same to within 3~Myr (due to the presences of
Wolf-Rayet features in all the complexes, as well derived from 
colour-colour diagrams, see Fig.~\ref{colour-f1}), then to pass
information of the formation of one to the next would require a speed
of $>~100$~km/s.  Typical sound speeds in GMCs are below
a few km/s (Dyson \& Williams 1997).  Thus, we see then that the
triggered formation 
from one complex to the next is not likely,  unless it was
triggered by older, presently unobserved, star formation.  One
possibility for the coeval nature of the present star formation is an
external perturbation, such as a large shock front, hit 
the parental GMC or GMC association at a very similar time throughout
the entire cloud.

\begin{figure}[!hbt]
\begin{centering}
      \caption{The continuum subtracted H$\alpha$ WFPC2 image and contours.  
      Note the large structure surrounding Complexes~4, 5, and~6. The
      shared envelope of excited gas surrounding these three
      complexes, along with the shared Wolf-Rayet features (see
      Fig.~\ref{image-vlt-wr})  may suggest a common age and a physical
      association.  The reconstructed IFU image of H$\beta$ intensity
      shows similar structure as the image shown here.} 
         \label{halpha-contours-f1}
	 \end{centering}
\end{figure}

\subsection{Relation to giant molecular clouds}

\begin{figure}[!hbt]
     \includegraphics[angle=0,width=8cm]{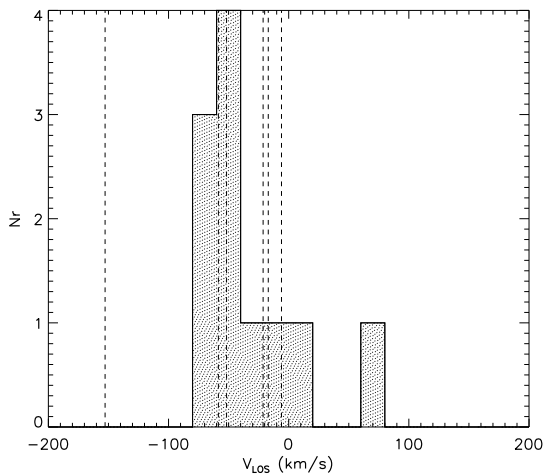}
      \caption{Histogram of the line of sight velocities of GMCs in
      NGC 4038/39 in the contact region ($\Delta 21''$ in RA and
      $\Delta 40''$ in Dec centered on Field~1, taken from Wilson et al.~(2003)).  The vertical dashed
      lines are the velocities of the complexes.  The furthest line to
      the left is the emission feature observed in the direction of
      Complex~3.} 
         \label{vel-gmcs}
\end{figure}

Figure~\ref{vel-gmcs} shows the giant molecular cloud velocities in
the overlap region of the NGC 4038/39 merger (Wilson et al.~2003).
The dashed vertical lines represent the velocities of the complexes in
Field~1.  The match is extremely good, showing that the young
complexes and the GMCs share the same general velocity distribution
within this region.  The main exception is the complex at $V=-153$~km/s
(with respect to the center of NGC 4038).  This is the emission
feature found in Complex~3.  Near-IR spectroscopy of the two 
members (Gilbert 2002) of Complex~3, show a continuum with no emission
features.  In particular the Br$\gamma$ emission feature is absent,
whereas it is expected to be present since we see relatively strong
H$\beta$ and H$\gamma$ emission.  Additionally, high resolution
spectroscopy centered on the Complex~3 shows some hint of a shifted emission
line $\sim 4$ arcseconds from the center of the slit (Sabine Mengel,
Priv. Comm.).  The nature of the emission feature is unknown.

Upon closer examination, however, the complexes in Field~1 are located
in a region devoid of molecular gas (see Fig.~4 in Wilson et
al.~2003), although the detection limit of their survey was rather
high at $\sim5\times10^{6} M_{\odot}$.  Thus they conclude that
regions which have no detectable CO emission but contain young star
clusters, either formed the clusters from lower masses clouds or the
clusters present have destroyed their parent GMCs. 

As shown in \S~\ref{agesphot} the complexes in Field~1 have luminous
masses between 
$\sim0.6-5 \times 10^{6} M_{\odot}$, suggesting that their
progenitor GMCs were greater than $5\times 10^{6} M_{\odot}$.  Thus,
our data are consistent with the idea that the young clusters and
complexes have destroyed their parent GMCs through their supernovae,
stellar winds and ionizing photons.

\section{Stability of the complexes}
\label{stability}

A number of recent studies (e.g. Kroupa 1998, Fellhauer \& Kroupa 2002) have suggested that
the clusters within complexes, such as the ones studied here, will
merge to form a single large star cluster in the center of the
complex.  This is expected to happen on timescales of a few $\times
10^{7}$ yrs (Kroupa 1998).  We therefore address the stability of the
complexes in this section.  We concentrate this analysis on Field~1,
where the complicated dynamics that occurs within the center of merging
galaxies should be less of a concern.

As a simple initial assumption, we estimate the tidal radius of
structures within a rotating disk.  This can be estimated by
$$r_{t} = (\frac{G M_{\rm complex}}{2 \cdot V_{G}^2})^{1/3} {\rm
R_{\rm G}^{2/3}} $$
where $M_{\rm complex}$, $V_{G}$, and $R_{G}$ are the mass of the complex, the
circular velocity of the galaxy at that point, and the distance to the
galactic center.  Assuming a rotational
velocity of 90 km/s (from Fig.~15 in Zhang et al.~2001) and a distance
between 2 and 4 kpc from the galactic center, we see that the tidal
radii of complexes with masses between $10^{5}$ and $10^{6} M_{\odot}$
are between 50 and 150 pc.  Additionally, given the steep velocity
gradient at the position of Complexes~1, 4, and 5 the approximation
given above will tend to overestimate the tidal radius.  Comparison
with Table~\ref{age:table} shows 
that most of the complexes are larger than their tidal radii, implying
that a significant fraction of each complex is unbound.

Also, due to the rapid removal of gas from these systems (see
\S~\ref{expansion}) a substantial fraction of the mass of the complex,
i.e. the gas left over from the star formation process, will be
removed in a very short time.  This will add to the amount of mass
lost by the complexes.  The inner regions of these complexes, however,
may be less affected by these processes due to the higher cluster
densities there, and merge to form a single massive object.  

Due to the expected loss of clusters from these complexes,
we expect them to contribute a large amount of their star
clusters into the surrounding field.  If merging does occur within the
centers of the complexes, we expect that the resulting object will
only contain a modest fraction of the total stellar mass of the
progenitor complex.

\section{Conclusions}
\label{conclusions}

We have presented the first results from a {\it VLT-VIMOS/IFU} survey
of merging galaxies.  Our targets were two fields within the Antennae
galaxies, one in the overlap region and the other near the center of
NGC 4038.  In particular we have studied six star cluster complexes
in the first region and two in the second region.  We have estimated
the extinction, metallicity, and velocity for each of these complexes
using the emission lines H$\gamma$, H$\beta$,
[OIII]$\lambda\lambda4959,5007$, and He~I ($\lambda$ 5876).  The
derived metallicities of the complexes are all above 0.5$Z_{\odot}$.   

Assuming a common extinction value for all sources in each complex
(see caveats to this assumption in \S~\ref{agesphot}), we are able to
correct the {\it HST} photometry of the individual sources within each
complex.  The colours of the star clusters (once corrected for
extinction) match the SSP models of Maraston (2005) extremely well.
We determined the ages of the clusters using these colour-colour diagrams. 

Complexes~4, 5, and 6 appear to be part of a larger structure
approximately (in projection) 550~pc in length.  All
three of these complexes show 
strong Wolf-Rayet features in their spectra, suggesting a common age
($\Delta({\rm age}) < 4$~Myr).  This grouping of complexes is most notably visible in
H$\alpha$, where the complexes seem to share a common ionized
envelope.  This common ionized envelope, as well as the similarity in
ages and metallicities of Complexes~4, 5, and 6 lead us to conclude that
they were part of a single GMC or GMC complex, which fragmented into
individual cluster complexes (e.g. a GMC association with multiple
massive cores). The H$\alpha$
spatial distributions are also 
hints that this collection of three
complexes is destroying the surrounding molecular gas and dust.  This
is presumably what has happened in the eastern section of
Field~1 where little gas and dust is seen.

The size of the large collection of complexes is approximately the
same as the scale at which the two-point correlation function of young
clusters drastically drops (i.e. at $\sim~735$~pc in H$\alpha$).  Thus, these
groupings of complexes are
likely to be the largest scale of correlated cluster formation.  This
scale corresponds to the size of large GMCs or GMC complexes, suggesting that a
single GMC can form multiple complexes, each in turn containing many
star clusters.  Alternatively, the velocity differences between the
complexes in this grouping ($\sim65$~km/s) may be an indication of
high-velocity GMC collisions as the primary triggering mechanism,
although the velocity difference does seem to follow the general
velocity distribution in this region of the galaxy.

The youngest complexes in Field~1 ($<$10~Myr) show rather small
velocity dispersions (20--40~km/s) as measured through the H$\beta$
and [OIII]$\lambda\lambda4959,5007$\AA~emission lines.  These values
can be understood as an expanding bubble, caused by the ionization and
winds from massive stars and supernovae, that is expanding into a
somewhat dense environment, i.e. the remnants of the progenitor GMC.

At least one complex in Field~1, Complex~2, shows evidence for a
multiple aged population.  The brightest cluster within the complex is
not spatially coincident with the H$\alpha$ emission in the region,
and has a photometric age estimate of $\sim7-15$ Myr.  Another bright
cluster within the complex has a younger photometric age estimate, and
is centered on the HII region.  Thus we see that even a single complex
can undergo multiple star formation episodes within
$\sim15$~Myr. Complex~Nuc~2 may also have a multiple-age cluster population,
although the clusters lie parallel to the extinction vector (in
colour-colour space) which suggests that they have similar ages but
that there is variable amounts of extinction within the complex.

Using the dereddened H$\beta$ emission line, we have estimated the
star formation rates within the complexes.  By normalizing the star
formation rates by their areas, we see that these complexes, with star
formation rates from 0.2 to 1.4 $M_{\odot}$/yr are
comparable to star forming regions in starburst galaxies.  Thus we
label these complexes as {\it localized starbursts}.  This coherent
and intense form of star formation may be responsible for the patchy
distribution of metals observed in the ISM of the Antennae (Fabbiano
et al.~2004).

The complexes in Field~1 have sizes similar to, or larger than, their
tidal radii, estimated by assuming a smooth circular velocity profile of the
galaxy.  (Our tidal radii estimates are an upper limit, as the velocity
profile of the galaxy, at the position of the complexes, is quite
irregular.)  Therefore we expect that these complexes will loose a
significant amount of their mass (individual stars and clusters) into
their surroundings.  The central regions of the complexes, however,
may be dense enough to allow a significant amount of merging between
clusters, as predicted by Kroupa (1998).

\begin{table*}[!p]
{\scriptsize
\parbox[b]{12cm}{
\centering
\caption[]{Positional and general information on the observed complexes.}
\begin{tabular}{c l l c c c c c c}
\hline\hline
\noalign{\smallskip}
ID$^{a}$   &      RA (J2000)       &       DEC (J2000) &
           $M_{V}^{a}$     & $A_{V}$   & Z  & V$_{r}^{b}$ &
           $\sigma^{c}$ &
            Notes\\
           &                       &                   &
           &               & (Z$_{\odot}$) & (km/s) &(km/s)  & \\
\hline
\\
1$^{d}$ & 12:01:55.60 & -18:52:21.8 & -13.8 & 0.0 & 0.97 & -58.5$\pm$ 
13.8  & 43 &weak WR features, contamination by close HII region\\
2 & 12:01:55.73 & -18:52:13.3 & -13.6 & 0.0 & 1.3 & -21.3$\pm$ 4.25 & 33
&\\
3a & 12:01:55.99 & -18:52:10.4 & -13.6 & 0.0  & 0.45 & -152.8$\pm$ 9.1 &  & \\
3b & & & & & & -17.0 $\pm$   19.2 &  73 &broad absorption feature superimposed on emission lines \\
4 & 12:01:54.86 & -18:52:12.9 & -13.5 & 0.72 & 0.75 & -51.7$\pm$ 7.5 &
37 & strong WR features\\
5 & 12:01:54.71 & -18:52:10.6 & -14.0 & 0.67 & 0.87 & -6.0$\pm$ 6.1 & 29 &strong WR features\\
6 & 12:01:54.58 & -18:52:07.2 & -13.1 & 0.91 & 0.76 & -3.5$\pm$ 5.7& 29 &strong WR features\\
1 Nuc & 12:01:52.97 & -18:52:08.4 & -14.6 & 0.0 & $> 2^{e}$ &
13.6$\pm$ 3.7 & --&strong WR features\\
2 Nuc & 12:01:53.04 & -18:52:02.7 & -14.3 & 1.0 & 1.67 & 60.5$\pm$
21.0 & -- &WR features\\

\noalign{\smallskip}

\noalign{\smallskip}
\hline
\end{tabular}
\begin{list}{}{}
\item[$^{\mathrm{a}}$] Within the 15 pixel (140 pc) radius used for source detection, uncorrected for extinction.
\item[$^{\mathrm{b}}$] Relative to the velocity of NGC 4038, taken to be V=1642 km/s.
\item[$^{\mathrm{c}}$] The velocity dispersion of the gas, as measured
from the H$\beta$ and O[III] lines.
\item[$^{\mathrm{d}}$] Measured extinction (A$_{V}$ = 1.3 mag) and metallicity presumably refer to the contaminating HII region nearby, as these quantities are measured from emission lines.  For the photometric analysis, no extinction correction has been applied.
\item[$^{\mathrm{e}}$] R$_{3}$ is out of the calibrated range.
\end{list}
\label{table:2info}

}
}
\end{table*}

\begin{table*}[tbh]
{\scriptsize
\parbox[b]{10cm}{
\begin{center}
\caption[]{Derived properties of the complexes}
\begin{tabular}{c c c c c c c c c }
\hline\hline
\noalign{\smallskip}
ID  & Field &  Age$^{a}$  & Age$^{b}$ & Mass$^{c}$ & M$_{\rm F555W}^{d}$ & R$_{\rm
F555W}$$^{e}$ & R$_{{\rm H}\alpha}$$^{e}$ & SFR$^{f}$\\
    & &  (Myr) & (Myr) & ($10^{6} M_{\odot}$) &   &  (pc) & (pc) & (M$_{\odot}$/yr) \\
\hline
\\
1     & 1 &  20--30   & -- & 2.0--2.5 & -11.50 & -- & -- & 1.40\\
2     & 1 &  6--16   & -- &  0.4--1.3 & -11.23 & 198 & 94 & 0.20\\
3     & 1 & 7--14 & -- & 0.4--1.2 & -12.31 & 200 & -- & -- \\

4     & 1 &  3.5--5.5  & 3-4.5 &  0.7--1.0 & -12.73 &  136 & 180 & 0.43\\
5     & 1 & 3--4   & 2.5-3 &	  1.1--1.2 & -13.15 & 175 & 189 & 0.35 \\
6     & 1 & 1--2  &  2.5-3 &  0.6--1.2  & -13.03    & 90    &  100   &
--\\

1 Nuc & 2 &  4--6    & 3.5-4.5 &	  0.9 & -13.49 & 123 & 238 & 0.47\\
2 Nuc &	2 & 3.5   & 3.5-4.5 &	  1.1 & -12.12 & -- & -- & 0.70\\

\noalign{\smallskip}

\noalign{\smallskip}
\hline
\end{tabular}
\begin{list}{}{}
\item[$^{\mathrm{a}}$] The age determined from the comparison of the
photometry of the brightest cluster to that of SSP models of solar metallicity.
\item[$^{\mathrm{b}}$] The age determined from Wolf-Rayet emission features
\item[$^{\mathrm{c}}$] The mass of the complex, corrected for
extinction, using the M/L ratio of the Maraston (2005) SSPs (solar
metallicity) and the age derived from the colours of the  brightest
star cluster in the complex.
\item[$^{\mathrm{d}}$] The magnitude of the brightest cluster within
the complex, corrected for extinction.
\item[$^{\mathrm{e}}$] The size (diameter) of the complex as measured in the
F555W and the H$\alpha$ filters.
\item[$^{\mathrm{f}}$] Based on the H$\beta$ emission flux.

\end{list}
\label{age:table}
\end{center}

}
}
\end{table*}

\begin{acknowledgements}
This paper has used observations taken from the HST data archive,
and has benefited from the NASA Extragalactic Database.  We kindly
thank Michele Cappellari for help implementing the {\it PPXF}
technique and Mark Gieles, Yuri Efremov \& Linda Smith
for useful discussions.  We also thank Henny Lamers for detailed
comments on earlier versions of the manuscript as well as for useful
discussions.  Finally, we thank the anonymous referee for helpful
comments and suggestions.
	
\end{acknowledgements}

\bibliographystyle{alpha}
\bibliography{../bib/astroph.bib,../bib/phd.bib,../bib/mark.bib}

\end{document}